\documentclass[prd,twocolumn,showpacs,floatfix,amsmath,amssymb,floatfix]{revtex4}
\usepackage{graphicx,color,dcolumn,booktabs,bm}
\usepackage{longtable,lscape}
\usepackage{txfonts}
\usepackage{overpic}
\usepackage{amssymb}
\usepackage{indentfirst}
\usepackage{feynmf}   %{feynmp}
\usepackage{slashed}  %for Feynman symbols
\usepackage{cases}
\usepackage{color}
\usepackage{multirow}
\usepackage{graphicx,color,dcolumn,booktabs,bm}
\usepackage{epstopdf}
\def\lrpartial{\buildrel\leftrightarrow\over\partial}
\graphicspath{{Figures/}} %

\begin{document}
%\begin{CJK}{GBK}{}

\title{Explaining the anomalous $\Upsilon(5S)\to \chi_{bJ}\omega$ decays through the hadronic loop effect}
\author{Dian-Yong Chen$^{1,3}$}\email{chendy@impcas.ac.cn}
\author{Xiang Liu$^{2,3}$\footnote{Corresponding author}}\email{xiangliu@lzu.edu.cn}
\author{Takayuki  Matsuki$^{4,5}$}\email{matsuki@tokyo-kasei.ac.jp}
\affiliation{
$^1$Theoretical Physics Division, Institute of Modern Physics of CAS, Lanzhou 730000, China\\
$^2$School of Physical Science and Technology, Lanzhou University, Lanzhou 730000, China\\
$^3$Research Center for Hadron and CSR Physics, Lanzhou University
$\&$ Institute of Modern Physics of CAS, Lanzhou 730000,
China\\
$^4$Tokyo Kasei University, 1-18-1 Kaga, Itabashi, Tokyo 173-8602, Japan\\
$^5$Theoretical Research Division, Nishina Center, RIKEN, Saitama 351-0198, Japan
}

\begin{abstract}
In this work, we carry out the study on $\Upsilon(5S)\to
\chi_{bJ}\omega$ ($J=0,1,2$) by considering the hadronic loop
mechanism. Our results show that the Belle's preliminary data of the
branching ratios for $\Upsilon(5S)\to \chi_{bJ}\omega$ can be well
reproduced in our calculation with a common parameter range, which
reflects the similarity among these $\Upsilon(5S)\to
\chi_{bJ}\omega$ decays of concern. %In addition, we also predict the
%branching ratios of $\Upsilon(5S)\to \chi_{b1}\eta$, which can reach
%up to $(0.80 \sim 1.67) \times 10^{-3}$. This prediction of
%considerable large value makes us suggest for Belle and forthcoming
%BelleII to search for $\Upsilon(5S)\to \chi_{b1}\eta$ in near
%future.

\end{abstract}

 \maketitle

%\section{Introduction}\label{sec1}

In the past years, the Belle Collaboration has reported some novel
phenomena relevant to the hidden bottom decays of $\Upsilon(5S)$. In
Ref. \cite{Abe:2007tk}, Belle indicated that the partial decay widths
of $\Upsilon(5S)\to \Upsilon(nS)\pi^+\pi^-$ are $10^2$ times larger than
those of $\Upsilon(mS)\to \Upsilon(nS)\pi^+\pi^-$, where $n,m=1,2,3$
and $m>n$, which is the puzzle in the $\Upsilon(5S)$
hidden-bottom dipion decays. There are two possible explanations for
this puzzle. One is that this large decay width can result
from the rescattering mechanism, where the hadronic loop composed of
the charmed mesons plays an important role \cite{Meng:2007tk,
Meng:2008dd}. Another possibility is that there is a tetraquark
state $Y_b$ near $\Upsilon(5S)$. According to this assumption, Ali {\it et al.} also
studied the $\pi^+\pi^-$ invariant mass spectrum and the $\cos\theta$
distribution of $Y_b\to \Upsilon(1S,2S)\pi^+\pi^-$. They claimed that
the experimental data can be well described under this explanation.
However, as indicated in Ref. \cite{Ali:2009es}, the result of
$Y_b\to \Upsilon(2S)\pi^+\pi^-$ is not consistent with the
corresponding experimental data. That is, in their calculation of
$Y_b\to \Upsilon(2S)\pi^+\pi^-$, they can describe $\pi^+\pi^-$ data.
If taking the same parameters to produce the $\cos\theta$
distribution, however, we found that the obtained $\cos\theta$ distribution
cannot fit the experimental data. Furthermore, in Ref.
\cite{Chen:2011qx}, the authors also studied $\Upsilon(5S)\to
\Upsilon(1S,2S)\pi^+\pi^-$ by the rescattering mechanism, where the
interference effect was considered. They met the same problem
when fitting the experimental data of $\Upsilon(5S)\to
\Upsilon(2S)\pi^+\pi^-$. Thus, a new puzzle was proposed in Ref.
\cite{Chen:2011qx}. Later, two charged bottomonium-like structures
$Z_b(10610)$ and $Z_b(10650)$ were reported by Belle
\cite{Belle:2011aa}, which also stimulated the authors in Ref.
\cite{Chen:2011qx} to find the relation between the observed $Z_b$
structures and the solution to this new puzzle. If introducing the
intermediate $Z_b$ contributions in $\Upsilon(5S)\to
\Upsilon(2S)\pi^+\pi^-$, the new puzzle mentioned above can be nicely solved,
which also results in the observation of the initial single
pion emission mechanism in Ref. \cite{Chen:2011pv} to explain why
there are two charged $Z_b$ structures near the $B\bar{B}^*$ and
$B^*\bar{B}^*$ thresholds. More theoretical predictions of
charged charmounium-like structures around the $D\bar{D}^*$ and
$D^*\bar{D}^*$ threshold were, of course, given in Ref. \cite{Chen:2011xk}.

The studies cited above show that the hadronic loop mechanism, as an
important non-perturbative QCD effect,  is indeed important to
$\Upsilon(5S)$ decays. Before applying the hadronic loop mechanism to
study the $\Upsilon(5S)$ decays, this mechanism was extensively
applied to study the decays of the higher bottomonium and
charmonium in Refs. \cite{Meng:2008bq, Chen:2009ah, Liu:2009vv,
Chen:2012nva, Chen:2013cpa, Zhao:2006gw} and achieved great
successes.

Very recently, Belle announced their observation of $\Upsilon(5S)\to
\chi_{bJ}\omega$ ($J=0,1,2$), which indicates that the
$\Upsilon(5S)\to \chi_{bJ}\omega$ decays also have large decay
widths; i.e., the measured branch ratios of $\Upsilon(5S)\to
\chi_{bJ}\omega$ are $<3.4\times 10^{-3}$,
$(1.64\pm0.23^{+0.30}_{-0.22})\times 10^{-3}$, and
$(0.57\pm0.22\pm0.07)\times 10^{-3}$ with $J=0,1,2$, respectively
\cite{Shen:2014rep, He:2014sqj}. It should be noticed that even though the tree-level contributions to $\Upsilon(5S)\to \chi_{bJ}\omega$ ($J=0,1,2$)
should be strongly suppressed due to the Okubo-Zweig-Iizuka
rule, such large decay widths are observed, which  again inspires
our interest in understanding such quantities. In this work, we
propose that the contribution from the hadronic loop should be
considered in studying $\Upsilon(5S)\to \chi_{bJ}\omega$. To give a
quantitative answer, we perform the concrete calculation, which is illustrated in the following. This investigation can, of course,
provide a good test of the hadronic loop mechanism.

%This paper is organized as follows. After the introduction, we
%present the formalism of the calculation of $\Upsilon(5S)\to
%\chi_{bJ}\omega$ by introducing the hadronic loop effect. In Sec.
%\ref{sec3}, the numerical result will be presented. The last section
%is the discussion and conclusion.

%\section{$\Upsilon(5S)$ decays into $\chi_{bJ}\omega$ with the hadronic
%loop contribution}\label{sec2}

\begin{figure}[htb]
\centering %
\scalebox{0.45}{\includegraphics{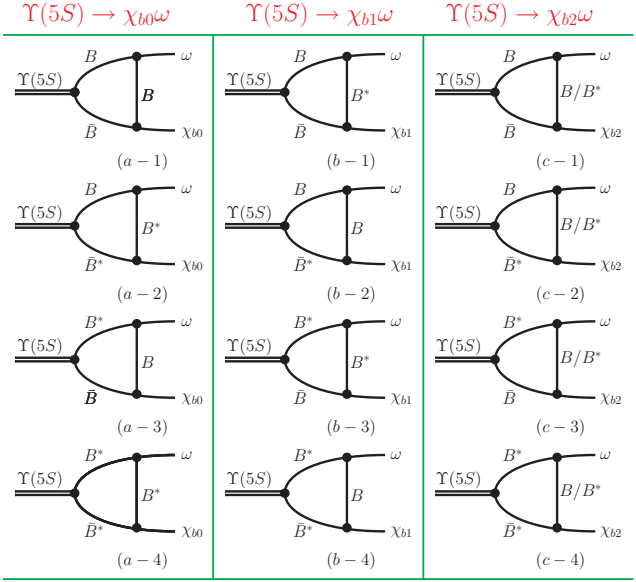}}
\caption{The necessary diagrams depicting $\Upsilon(5S) \to
\chi_{bJ} \omega$ decays under the hadronic loop effect. \label{Fig:Chib2J}}
\end{figure}

\begin{figure}[htb]
\centering %
\scalebox{0.7}{\includegraphics{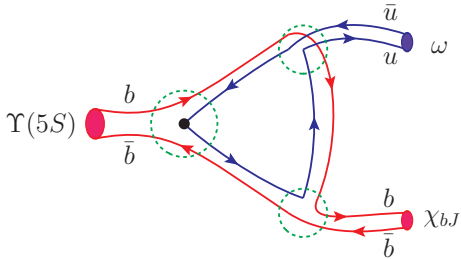}}
\caption{The quark level diagram depicting $\Upsilon(5S) \to
\chi_{bJ} \omega$ decay under the hadronic loop effect. \label{Fig:quark}}
\end{figure}

$\Upsilon(5S)$ as a higher bottomonium is above the threshold of a
pair of bottom mesons, where $\Upsilon(5S)$ mainly decays into $B^{(\ast)}
\bar{B}^{(\ast)}$, which means that there exists the strong coupling between
$\Upsilon(5S)$ and a bottom meson pair. Thus, the hadronic loop effect can play an important role in
the decay of $\Upsilon(5S)$, as just briefly reviewed above. Under the hadronic loop mechanism, these discussed $\Upsilon(5S)\to
\chi_{bJ}\omega$ processes occur via the intermediate $B^{(*)}$ meson loops. In Fig. \ref{Fig:Chib2J}, the diagrams describing the $\Upsilon(5S)\to \chi_{bJ}\omega$ decays are given, where an intermediate bottom meson pair can
transit into final states  $\chi_{bJ}\omega$ by exchanging a proper bottom meson.
Instead of the hadronic description for $\Upsilon(5S)\to \chi_{bJ}\omega$, we can give a quark level description of the hadronic loop contribution in Fig. \ref{Fig:quark}. Here, a fermion line in red denotes bottom quark while a blue line corresponds to the light quark. $\Upsilon(5S)$ first dissolves into two virtual bottom mesons and then this bottom meson pair can turn into $\chi_{bJ}\omega$ via an exchange of an appropriate bottom meson. The matrix element of $\Upsilon(5S)\to \chi_{bJ}\omega$ via hadronic loop effect can be depicted as
\begin{eqnarray}
\mathcal{M}(\Upsilon(5S)\to \chi_{bJ}\omega)=\sum_j\langle
\chi_{bJ}\omega|\mathcal{H}_2|j\rangle\langle j|\mathcal{H}_1|\Upsilon(5S).
\end{eqnarray}
The corresponding description at the hadron level is listed in Fig. \ref{Fig:Chib2J}.

In the heavy quark limit, the wave function of a heavy-light meson is independent of the flavor and spin of the heavy quark; therefore, this wave function can be characterized by the angular momentum of the light degrees of freedom, which is $\vec s_\ell=\vec s_q +\vec \ell$. Each value of $s_\ell=|\vec s_\ell|$ corresponds to a degenerate doublet of states with the total angular momentum $J=s_\ell \pm 1/2$. For the bottom meson with $\ell=0$, the doublet formed by the bottom pseudoscalar and vector meson is represented in \cite{Kaymakcalan:1983qq, Oh2000qr, Casalbuoni1996pg,
Colangelo2002mj},
\begin{eqnarray}
H^{(Q\bar{q})}=\frac{1+ \slashed{v}}{2} \left[\mathcal{B}^\ast_\mu \gamma^\mu-\mathcal{B} \gamma^5\right].
\end{eqnarray}

For the heavy quarkonium, the degeneracy is expected under the rotations of the two heavy quark spins, although the heavy quark flavor symmetry does not hold any more. This allows heavy quarkonium with the same angular momentum $\ell$ to form a multiplet. For the bottomonia with $\ell=0$, $\eta_b$ and $\Upsilon$ form a doublet in the form
\begin{eqnarray}
R^{(Q\bar{Q})}=\frac{1+\slashed{v}}{2} \left[\Upsilon^\mu \gamma_\mu -\eta_b \gamma_5\right]\frac{1-\slashed{v}}{2}.
\end{eqnarray}
In a similar way, a spin multiplet corresponding to the $P$-wave bottomonia is,
\begin{eqnarray}
&&{P^{(Q\bar{Q})}}^\mu =\frac{1+\slashed{v}}{2} \Big[\chi_{b2}^{\mu \alpha} \gamma_\alpha +\frac{1}{\sqrt{2}}  \varepsilon^{\mu \alpha \beta \gamma} v_\alpha \gamma_\beta \chi_{b1 \gamma} \nonumber\\
&&\hspace{20mm}+\frac{1}{\sqrt{3}} \big(\gamma^\mu -v^\mu \big) \chi_{b0} +h_b^\mu \gamma_5 \Big] \frac{1-\slashed{v}}{2}.
\end{eqnarray}

With these multiplets, we can construct the general form of the coupling between heavy quarkonium and heavy meson. The related effective Lagrangians involved in the present work are \cite{Casalbuoni1996pg}
\begin{eqnarray}
\mathcal{L}_s&=& ig \mathrm{Tr} \left[R^{(Q\bar{Q})} \bar{H}^{(\bar{Q}q)} \gamma^\mu \lrpartial_\mu \bar{H}^{(Q\bar{q})} \right] +H.c., \nonumber\\
\mathcal{L}_p &=& ig_1 \mathrm{Tr} \left[P^{(Q\bar{Q}) \mu} \bar{H}^{(\bar{Q}q)} \gamma_\mu \bar{H}^{(Q\bar{q})}\right] +H.c.,
\end{eqnarray}
where $H^{(\bar{Q}q)}$ represents the heavy-light meson containing a heavy antiquark $\bar{Q}$, which can be obtained by applying the charge conjugation operation to $H^{(Q\bar{q})}$. Expanding the above Lagrangians, we can obtain the following effective couplings:
\begin{eqnarray}
\mathcal{L}_{\Upsilon(5S) \mathcal{B}^{(\ast)} \mathcal{B}^{(\ast)}}
&=& -ig_{\Upsilon(5S) \mathcal{BB} } \Upsilon_\mu (\partial^\mu
\mathcal{B} \mathcal{B}^\dagger- \mathcal{B}
\partial^\mu \mathcal{B}^\dagger) \nonumber\\ && + g_{\Upsilon(5S)
\mathcal{B}^\ast \mathcal{B}} \varepsilon^{\mu \nu \alpha \beta}
\partial_\mu \Upsilon_\nu (\mathcal{B}^\ast_\alpha \lrpartial_\beta
\mathcal{B}^\dagger -\mathcal{B} \lrpartial_\beta
\mathcal{B}_\alpha^{\ast \dagger} ) \nonumber\\ && + ig_{\Upsilon
\mathcal{B}^\ast \mathcal{B}^\ast} \Upsilon^\mu
(\mathcal{B}^\ast_\nu \partial^\nu \mathcal{B}^{\ast \dagger}_\mu
-\partial^\nu \mathcal{B}^{\ast}_\mu \mathcal{B}^{\ast \dagger}_\nu
-\mathcal{B}^\ast_\nu \lrpartial_\mu \mathcal{B}^{\ast \nu
\dagger}),\nonumber\\
\end{eqnarray}
\begin{eqnarray}
\mathcal{L}_{\chi_{bJ} \mathcal{B}^{(\ast)} \mathcal{B}^{(\ast)}}
&=& - g_{\chi_{b0} \mathcal{B} \mathcal{B} } \chi_{b0} \mathcal{B}
\mathcal{B}^\dagger - g_{\chi_{b0} \mathcal{B}^\ast
\mathcal{B}^\ast} \chi_{b0} \mathcal{B}_{\mu}^\ast \mathcal{B}^{\ast
\mu\dagger } \nonumber\\ &&  +i g_{\chi_{c1} \mathcal{B}
\mathcal{B}^\ast} \chi_{b1}^\mu ( \mathcal{B}^{\ast }_\mu
\mathcal{B}^\dagger - \mathcal{B} \mathcal{B}^{\ast \dagger}_\mu )
\nonumber\\ &&  - g_{\chi_{b2} \mathcal{B} \mathcal{B}}
\chi_{b2}^{\mu \nu}
\partial_\mu \mathcal{B} \partial_\nu \mathcal{B}^\dagger
+ g_{\chi_{b2} \mathcal{B}^\ast \mathcal{B}^\ast} \chi_{b2}^{\mu
\nu}
\mathcal{B}^\ast_{\mu} \mathcal{B}^{\ast \dagger}_\nu \nonumber\\
&&  -ig_{\chi_{b2} \mathcal{B}^\ast \mathcal{B}} \varepsilon_{\mu
\nu \alpha \beta} \partial^\alpha \chi_{b2}^{\mu \rho}
(\partial_\rho \mathcal{B}^{\ast \nu} \partial^\beta
\mathcal{B}^\dagger -\partial^\beta \mathcal{B}
\partial_\rho \mathcal{B}^{\ast \nu \dagger} ),\nonumber\\
\end{eqnarray}
\begin{eqnarray}
\mathcal{L}_{\mathcal{B}^{(\ast)}\mathcal{B}^{(\ast)} \mathcal{V}}
&=& -ig_{\mathcal{B} \mathcal{B}\mathcal{V}} \mathcal{B}_i^\dagger
\lrpartial^\mu \mathcal{B}^j (\mathcal{V}_\mu)^i_j -2
f_{\mathcal{B}^\ast \mathcal{B} \mathcal{V}} \varepsilon_{\mu \nu
\alpha \beta}  \nonumber\\&& \times (\partial^\mu
\mathcal{V}^\nu)^i_j (\mathcal{B}^\dagger_i \lrpartial^\alpha
\mathcal{B}^{\ast \beta j} -\mathcal{B}_i^{\ast \beta \dagger}
\lrpartial^\alpha \mathcal{B}^j) \nonumber\\&& +ig_{\mathcal{B}^\ast
\mathcal{B}^\ast \mathcal{V}} \mathcal{B}^{\ast \nu \dagger}_i
\lrpartial^\mu \mathcal{B}^{\ast j}_\nu (\mathcal{V}_\mu)^i_j
\nonumber\\ &&+4if_{\mathcal{B}^\ast \mathcal{B}^\ast \mathcal{V}}
\mathcal{B}^{\ast \dagger}_{i\mu} (\partial^\mu \mathcal{V}^\nu
-\partial^\nu \mathcal{V}^\mu)^i_j \mathcal{B}^{\ast j}_\nu,
\end{eqnarray}
where $\mathcal{V}$ is the matrix of the vector octet, which is in
the form
\begin{eqnarray}
\mathcal{V} &=&
 \left(
 \begin{array}{ccc}
\frac{1}{\sqrt{2}} (\rho^{0}+ \omega) & \rho^{+} & K^{*+}\\
\rho^{-} & \frac{1}{\sqrt{2}}(-\rho^0 +\omega) &  K^{*0}\\
 K^{*-} & \bar{K}^{*0} & \phi
 \end{array}
 \right),
\end{eqnarray}

With the  above effective Lagrangian, we can write out the
amplitudes of hadronic loop contributions to $\Upsilon(5S) \to
\chi_{bJ} \omega\,(J=0,1,2)$. For $\Upsilon(5S) \to \chi_{b0}
\omega $, the amplitudes corresponding to 
Fig. \ref{Fig:Chib2J} (a-1)-(a-4) are
\begin{eqnarray}
\mathcal{M}_{(a-1)} &=& \int \frac{d^q}{(2\pi)^4} \Big[ ig_{\Upsilon
BB} \epsilon_\Upsilon^\mu (ip_{1\mu}-ip_{2\mu}) \Big] \Big[
-ig_{BBV} \epsilon_{\omega}^\nu
(-ip_{1\nu}\nonumber\\&&-iq_\nu)\Big] \Big[- g_{\chi_{b0} BB} \Big]
\frac{1}{p_1^2-m_B^2} \frac{1}{p_2^2-m_B^2}
\frac{\mathcal{F}^2(\Lambda)}{q^2-m_B^2} , \label{Eq:Ma-1}
\end{eqnarray}
\begin{eqnarray}
\mathcal{M}_{(a-2)} &=& \int \frac{d^q}{(2\pi)^4} \Big[g_{\Upsilon
B^\ast B} \varepsilon_{\rho \mu \alpha \beta} (-ip_0^\rho)
\epsilon_\Upsilon^\mu (-ip_2^\beta
+ip_1^\beta)\Big]\nonumber\\&&\times \Big[-2 f_{B^\ast BV}
\varepsilon_{\lambda \nu \theta \phi} (ip_3^\lambda)
\epsilon_\omega^\nu (ip_1^\theta+iq^\theta) \Big] \Big[-
g_{\chi_{b0} B^\ast B^\ast}\Big]\nonumber\\&&\times
\frac{1}{p_1^2-m_B^2} \frac{-g^{\alpha \tau}+p_2^\alpha
p_2^\tau/m_{B^\ast}^2}{p_2^2-m_{B^\ast}^2}
\frac{-g^{\phi}_{\tau}+q^\phi q_\tau/m_{B^\ast}^2}{q^2-m_{B^\ast}^2}
,\nonumber\\&& \times \mathcal{F}^2(\Lambda)
\end{eqnarray}
\begin{eqnarray}
\mathcal{M}_{(a-3)} &=& \int \frac{d^q}{(2\pi)^4} \Big[g_{\Upsilon
B^\ast B} \varepsilon_{\rho \mu \alpha \beta} (-ip_0^\rho)
\epsilon_\Upsilon^\mu (ip_2^\beta -ip_1^\beta)\Big] \nonumber\\
\nonumber\\&&\times\Big[ -2 f_{B^\ast BV} \varepsilon_{\lambda \nu
\theta \phi} (ip_3^\lambda) \epsilon_\omega^\nu
(-ip_1^\theta-iq^\theta)\Big]
\Big[-g_{\chi_{b0}BB}\Big]  \nonumber\\
&&\times\frac{-g^{\alpha \phi}+p_1^\alpha
p_1^\phi/m_{B^\ast}^2}{p_1^2-m_{B^\ast}^2} \frac{1}{p_2^2-m_B^2}
\frac{1}{q^2-m_B^2} \mathcal{F}^2(\Lambda),
\end{eqnarray}
\begin{eqnarray}
\mathcal{M}_{(a-4)} &=&  \int \frac{d^q}{(2\pi)^4} \Big[ig_{\Upsilon
B^\ast B^\ast} \epsilon_\Upsilon^\mu (ip_{2\alpha}g_{\mu
\beta}-ip_{1\beta} g_{\mu \alpha}
-(ip_{2\mu}\nonumber\\&&-ip_{1\mu})g_{\alpha \beta})\Big]
\Big[ig_{B^\ast B^\ast V} (-ip_{1\nu}-iq_\nu)\epsilon_\omega^\nu
g_{\theta \phi} \nonumber\\&&+4i f_{B^\ast B^\ast V}
\epsilon_\omega^\nu (ip_{3\phi} g_{\nu \theta}-ip_{3\theta} g_{\nu
\phi}) \Big]\Big[- g_{\chi_{b0} B^\ast B^\ast}\Big]
\nonumber\\&&\times\frac{-g^{\alpha \theta}+ p_1^\alpha
p_1^\theta/m_{B^\ast}^2}{p_1^2-m_{B^\ast}^2} \frac{-g^{\beta \tau}+
p_2^\beta p_2^\tau/m_{B^\ast}^2}{p_2^2-m_{B^\ast}^2}
\nonumber\\
&&\times \frac{-g^{\phi}_{\tau}+q^\phi
q_\tau/m_{B^\ast}^2}{q^2-m_{B^\ast}^2} \mathcal{F}^2(\Lambda),
\end{eqnarray}
respectively. Similarly, we can write out the
amplitudes for $\Upsilon(5S) \to \chi_{b1} \omega$ corresponding to
Fig. \ref{Fig:Chib2J} (b-1)-(b-4), which are
\begin{eqnarray}
%%
%% BBbar BStar
%%
\mathcal{M}_{(b-1)} &=& \int \frac{d^q}{(2\pi)^4} \Big[ ig_{\Upsilon
BB} \epsilon_\Upsilon^\mu (ip_{1\mu}-ip_{2\mu}) \Big] \Big[-2
f_{BB^\ast V}  \nonumber\\&& \times \varepsilon_{\lambda \nu \alpha
\beta } (ip_3^\lambda) \epsilon_\omega^\nu
(ip_1^\alpha-iq^\alpha)\Big] \Big[ig_{\chi_{b1} B^\ast B}
\epsilon_{\chi{b1}}^\theta\Big] \nonumber\\ && \times
\frac{1}{p_1^2-m_B^2} \frac{1}{p_2^2-m_B^2} \frac{-g^\beta_\theta +
q^\beta q_\theta/m_{B^\ast}^2}{q^2
-m_{B^\ast}^2} \mathcal{F}^2(\Lambda),\nonumber\\
\end{eqnarray}
\begin{eqnarray}
%%
%% BBStarbar B
%%
\mathcal{M}_{(b-2)} &=& \int \frac{d^q}{(2\pi)^4} \Big[g_{\Upsilon
B^\ast B} \varepsilon_{\rho \mu \alpha \beta} (-ip_0^\rho)
\epsilon_\Upsilon^\mu (-ip_2^\beta +ip_1^\beta)\Big] \nonumber\\
\nonumber\\&& \times \Big[-ig_{BBV} (-ip_{1 \nu}-iq_{\nu})
\epsilon_\omega^\nu \Big] \Big[-ig_{\chi_{b1} B^\ast B}
\epsilon_{\chi_{b1}}^\theta \Big] \nonumber\\ && \times
\frac{1}{p_1^2-m_B^2} \frac{-g^{\alpha}_\theta +p_2^\alpha
p_{2\theta}/m_{B^\ast}^2}{p_2^2-m_{B^\ast}^2} \frac{1}{q^2
-m_{B}^2} \mathcal{F}^2(\Lambda),\nonumber\\
\end{eqnarray}
\begin{eqnarray}
%%
%% BStar Bbar BStar
%%
\mathcal{M}_{(b-3)} &=& \int \frac{d^q}{(2\pi)^4} \Big[g_{\Upsilon
B^\ast B} \varepsilon_{\rho \mu \lambda \phi} (-ip_0^\rho)
\epsilon_\Upsilon^\mu (ip_2^\phi -ip_1^\phi)\Big] \nonumber\\
\nonumber\\&&\times \Big[ ig_{B^\ast B^\ast V} (ip_{1\nu}-iq_\nu)
\epsilon_{\omega}^\nu g_{\alpha \beta} +4if_{B^\ast B^\ast V}
\nonumber\\&& \times (ip_{3 \beta} g_{\nu \alpha} -ip_{3 \alpha
}g_{\nu \beta}) \epsilon_\omega^\nu  \Big] \Big[ ig_{\chi_{b1}
B^\ast B} \epsilon_{\chi_{b1}}^\theta \Big] \nonumber\\ && \times
\frac{-g^{\lambda \alpha}+p_1^\lambda p_1^\alpha/m_{B^\ast}^2
}{p_1^2-m_{B^\ast}^2} \frac{1}{p_2^2-m_{B}^2} \nonumber\\ && \times
\frac{-g^{\beta}_{\theta}+q^\beta q_\theta/m_{B^\ast}^2}{q^2
-m_{B^\ast}^2} \mathcal{F}^2(\Lambda),
\end{eqnarray}
\begin{eqnarray}
%%
%% BStar BStarbar B
%%
\mathcal{M}_{(b-4)} &=& \int \frac{d^q}{(2\pi)^4} \Big[ig_{\Upsilon
B^\ast B^\ast} \epsilon_\Upsilon^\mu (ip_{2\alpha}g_{\mu
\beta}-ip_{1\beta} g_{\mu \alpha}
-(ip_{2\mu}\nonumber\\&&-ip_{1\mu})g_{\alpha \beta})\Big] \Big[-2
f_{B^\ast B V} \varepsilon_{\lambda \nu \kappa \phi} (ip_3^\lambda)
\epsilon_\omega^\nu (-ip_1^\kappa -iq^\kappa)\Big]\nonumber\\ &&
\times\Big[- ig_{\chi_{b1} B^\ast B} \epsilon_{\chi_{b1}}^\theta
\Big]  \frac{-g^{\alpha \phi}+p_1^\alpha p_1^\phi/m_{B^\ast}^2
}{p_1^2-m_{B^\ast}^2}\nonumber\\ && \times \frac{-g^{\beta}_{
\theta} +p_2^\beta p_{2\theta}/m_{B^\ast}^2}{p_2^2- m_{B^\ast}^2 }
\frac{1}{q^2 -m_{B}^2} \mathcal{F}^2(\Lambda). \label{Eq:Mb-4}
\end{eqnarray}

The hadronic loop contribution to $\Upsilon(5S) \to \chi_{b2} \omega$
is listed in Fig. \ref{Fig:Chib2J} (c-1)-(c-4). In these diagrams,
the exchanged bottom meson can be $B$ meson or $B^\ast$ meson. The
concrete amplitudes are collected in the Appendix.

Considering the isospin symmetry and charge symmetry, we obtain the total
amplitude of $\Upsilon(5S) \to \chi_{b0} \omega$
\begin{eqnarray}
\mathcal{M}^{\mathrm{Tot}}_{\Upsilon(5S)\to \chi_{bJ} \omega}
=4\sum_{j=1}^4 \mathcal{M}_{(i-j)},
\end{eqnarray}
where $i=a$, $b$, and $c$ correspond to $\Upsilon(5S) \to \chi_{b0}
\omega$, $\Upsilon(5S)\to \chi_{b1} \omega$, and $\Upsilon(5S)\to
\chi_{b2} \omega$, respectively. The amplitudes of
$\mathcal{M}_{(a-j)}$ and $\mathcal{M}_{(b-j)}$ have been presented in
Eqs. (\ref{Eq:Ma-1})-(\ref{Eq:Mb-4}). The amplitudes
$\mathcal{M}_{(c-i)}$ are defined as
$\mathcal{M}_{(c-i)}=\mathcal{M}_{(c-i)}^B
+\mathcal{M}_{(c-i)}^{B^\ast}$. With above amplitudes, the partial
decay width reads as
\begin{eqnarray}
\Gamma_{\Upsilon(5S) \to \chi_{bJ} \omega} = \frac{1}{24\pi}
\frac{|\vec{p}_\omega|}{m_{\Upsilon(5S)}^2}
|\overline{\mathcal{M}^{\mathrm{Tot}}_{\Upsilon(5S) \to \chi_{bJ}
\omega}}|^2,
\end{eqnarray}
where the overline indicates the sum over the polarization vectors of
$\Upsilon(5S)$ and $\omega$. In addition, we define
$|\vec{p}_\omega|=\lambda^{1/2}(m_{\Upsilon(5S)}^2, m_{\chi_{b0}}^2
,m_{\omega}^2)$ with the K$\ddot{\mathrm{a}}$llen function
$\lambda(x,y,z)=x^2+y^2+z^2-2xy-2xz-2yz$.

Adopting the similar approach, we can obtain the amplitudes of $\Upsilon(5S) \to
\chi_{b1}\omega$ and $\Upsilon(5S)\to\chi_{b2} \omega$, which correspond to
Fig. \ref{Fig:Chib2J}  (b-1)-(b-4) and Fig. \ref{Fig:Chib2J} (c-1)-(c-4), respectively.
In the amplitudes, we introduce a form factor in the monopole form to depict
the internal structures as well as the offshell effect of the
exchanged bottom mesons, where the form factor is taken as
$\mathcal{F}(\Lambda) =({m_{E}^2 -\Lambda^2})/({q^2-\Lambda^2})$, with $m_E$ the exchagend boson mass.
In the heavy quark limit, $B$ and $B^\ast$ are degenerate and the space
wave functions of $B$ and $B^\ast$ are the same. Thus, in the present
work, we parameterize the cutoff $\Lambda$ as
$\Lambda=(m_B+m_{B^\ast})/2 +\alpha_{\Lambda} \Lambda_{QCD}$ with
$\Lambda_{QCD}=0.22$ GeV.

\renewcommand{\arraystretch}{1.4}

\begin{table}[htpb]
\centering \caption{The coupling constants of $\Upsilon(5S)$ interacting with $B^{(\ast)} \bar{B}^{(\ast)}$. Here, we also list the corresponding branching ratios.\label{Tab:UpsilonCP}}
\begin{tabular}{cccccc}
  \toprule[1pt]
Final state  & $\mathcal{B}(\%)$ & Coupling &
Final state  & $\mathcal{B}(\%)$ & Coupling  \\
 \midrule[1pt] %
$B\bar{B}$                 & 5.5   & 1.77 &
$B\bar{B}^\ast$            & 13.7  & 0.14 GeV$^{-1}$ \\
$B^\ast \bar{B}^\ast $     & 38.1  & 2.25  \\
  \bottomrule[1pt]
\end{tabular}
\end{table}

Since $\Upsilon(5S)$ is above the threshold of $B^{(\ast)} \bar{B}^{(\ast)}$, the coupling constants between $\Upsilon(5S)$ and $B^{(\ast)}
\bar{B}^{(\ast)}$ can be evaluated by partial decay width of
$\Upsilon(5S) \to B^{(\ast)} \bar{B}^{(\ast)}$. The partial decay width and
the corresponding coupling constants are listed in Table
\ref{Tab:UpsilonCP}.
If a vector boson multiplet is included, the effective Lagrangian both with pseudoscalars and vector bosons is constructed as in Refs. \cite{Kaymakcalan:1983qq,Oh2000qr,Casalbuoni1996pg,Colangelo2002mj}. This Lagrangian includes only one gauge coupling $g_1$ in the heavy quark limit so that all of the coupling constants are related to this gauge coupling.
In the heavy quark limit, the coupling constants
of $\chi_{bJ} \mathcal{B}^{(\ast)} \mathcal{B}^{(\ast)}$ are related
to the gauge coupling $g_1$ by
\begin{eqnarray}
g_{\chi_{b0} \mathcal{BB}}&=&2\sqrt{3} g_1\sqrt{m_{\chi_{b0}}} m_{\mathcal{B}}, \ \
g_{\chi_{b0} \mathcal{B}^\ast \mathcal{B}^\ast } =\frac{2}{\sqrt{3}}
g_1 \sqrt{m_{\chi_{b0}}} m_{\mathcal{B}^\ast},\nonumber\\
g_{\chi_{b1} \mathcal{B}\mathcal{B}^\ast} &=& 2\sqrt{2} g_1
\sqrt{m_{\chi_{b1}} m_{\mathcal{B}}m_{\mathcal{B}^\ast}}, \ \
g_{\chi_{b2} \mathcal{BB}} =2g_1 \frac{\sqrt{m_{\chi_{b0}}}
}{m_{\mathcal{B}}}, \nonumber\\
g_{\chi_{b2} \mathcal{B} \mathcal{B}^\ast} &=& g_1
\sqrt{\frac{m_{\chi_{b2}}}{m_{\mathcal{B}^\ast}^3 m_{\mathcal{B}}}},
\ \ g_{\chi_{b2} \mathcal{B}^\ast \mathcal{B}^\ast}
=4g_1\sqrt{m_{\chi_{b2}}} m_{\mathcal{B}^\ast},\nonumber
\end{eqnarray}
where we take the gauge coupling $g_1=-\sqrt{m_{\chi_{b0}}\over
3}\frac{1}{f_{\chi_{b0}}}$ and $f_{\chi_{b0}}=175 \pm 55$ MeV is the
decay constant of $\chi_{b0}$ \cite{Veliev:2010gb}. The coupling
constants between light vector mesons and bottom
mesons are
\begin{eqnarray}
%%
%%
%% B(Star)B(Star)V
 g_{\mathcal{B}\mathcal{B} \mathcal{V}} &=&g_{\mathcal{B}^\ast
\mathcal{B}^\ast \mathcal{V}}=\frac{\beta g_V}{\sqrt{2}},\nonumber\\
f_{\mathcal{B} \mathcal{B}^\ast \mathcal{V}}
&=&\frac{f_{\mathcal{B}^\ast \mathcal{B}^\ast
\mathcal{V}}}{m_{\mathcal{B}^\ast}} =\frac{\lambda g_V}{\sqrt{2}},\nonumber
\end{eqnarray}
where the gauge coupling $\beta=0.9$, $\lambda=0.56 \
\mathrm{GeV}^{-1}$, and $g_V=m_\rho/f_\pi$ with pion decay constant
$f_\pi=132$ MeV \cite{Cheng:1992xi, Yan:1992gz, Wise:1992hn,
Burdman:1992gh}.

\begin{figure}[htb]
\centering %
\scalebox{1}{\includegraphics{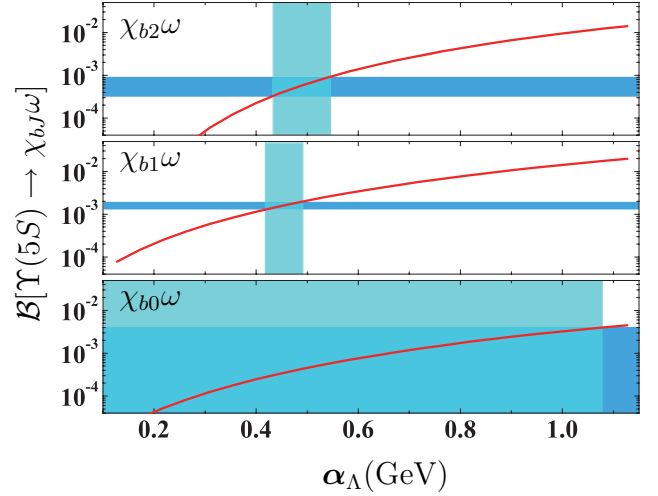}} %
\caption{The branching ratios of $\Upsilon(5S) \to \chi_{bJ} \omega$
dependent on the parameter $\alpha_\Lambda$. The horizontal bands are
the experimental data measured by the Belle Collaboration, while the
vertical bands indicate the $\alpha_\Lambda$ range when our results overlap with the Belle
data. \label{Fig:omega}}
\end{figure}

With above preparations, we can evaluate the hadronic loop
contributions to $\Upsilon(5S) \to \chi_{bJ} \omega$ decays. The
$\alpha_\Lambda$ is introduced as a free parameter in the cutoff
$\Lambda$ of the form factor. This parameter is usually dependent on
particular process and taken to be of the order of unity. In Fig.
\ref{Fig:omega}, we present the $\alpha_\Lambda$ dependence of the
branching ratio of $\Upsilon(5S) \to \chi_{bJ} \omega$. The
experimental data from the Belle Collaboration \cite{Shen:2014rep,He:2014sqj}
are also presented in comparison with our calculated results.

From Fig. \ref{Fig:omega}, we notice that our theoretical estimate
can reproduce the experimental data given by the Belle Collaboration
\cite{Shen:2014rep,He:2014sqj}. For $\Upsilon(5S) \to\chi_{b0} \omega$, only
the upper limit was give by the experimental measurement, which is
$\mathcal{B}(\Upsilon(5S) \to \chi_{b0} \omega)<3.4\times 10^{-3}$,
where our result overlaps with the experimental data when taking the
range $\alpha_\Lambda<1.09$. As for the discussed $\Upsilon(5S) \to
\chi_{b1} \omega$ and $\Upsilon(5S) \to \chi_{b2} \omega$ decays,
our calculation can be fitted to the corresponding experimental
values when taking $0.41<\alpha_\Lambda<0.48$ and
$0.43<\alpha_\Lambda <0.54$, respectively. Moreover, we need to
emphasize that there exists a common $\alpha_\Lambda$ range
$0.43<\alpha_\Lambda<0.48$ for all $\Upsilon(5S) \to \chi_{bJ}
\omega$ decays, which reflects the similarity among these three
decays. With this common $\alpha_\Lambda$ range, we can further
restrict the branching ratio of $\Upsilon(5S) \to \chi_{b0} \omega$,
which is $3.00 \times 10^{-4}<\mathcal{B}(\Upsilon(5S) \to \chi_{b0}
\omega)<4.05 \times 10^{-4}$, where this branching ratio is about
1 order smaller than the corresponding upper limit reported by
Belle \cite{Shen:2014rep,He:2014sqj} , which can be tested in a future
experiment.

In summary, being stimulated by the recent preliminary results of
$\Upsilon(5S)\to \chi_{bJ}\omega$ released by Belle
\cite{Shen:2014rep,He:2014sqj}, we have studied the $\Upsilon(5S)\to
\chi_{bJ}\omega$ decays through the hadronic loop mechanism. In the
past years, there were some experimental
\cite{Abe:2007tk,Belle:2011aa} and theoretical progresses
\cite{Meng:2007tk,Meng:2008dd,Chen:2011pv,Chen:2011xk} on the
$\Upsilon(5S)$ decays, which show that the hadronic loop mechanism
can be an important effect on the $\Upsilon(5S)$ decays. The present
investigation provides a further test of  the hadronic loop effect.
Our calculation indicates that the Belle data of $\Upsilon(5S)\to
\chi_{bJ}\omega$ can be reproduced when the hadronic loop mechanism
is considered in $\Upsilon(5S)\to \chi_{bJ}\omega$. What is more
important is that there exists a common $\alpha_\Lambda$ range for
all $\Upsilon(5S)\to \chi_{bJ}\omega$ decays, which is due to the
similarity among $\Upsilon(5S)\to \chi_{bJ}\omega$ with $J=0,1,2$.
In addition, we further constrain the branching ratio of
$\Upsilon(5S)\to \chi_{b0}\omega$ by the obtained common parameter
range, which can be tested in future experiments.

\section*{APPDNDIX:  THE DECAY AMPLITUDES OF $\Upsilon(5S) \to \chi_{b2} \omega$}

We collected the $\Upsilon(5S) \to \chi_{b2} \omega$ decay amplitudes, i.e.,

\begin{eqnarray}
%%
%% B Bbar B
%%
\mathcal{M}_{(c-1)}^{B} &=& \int \frac{d^q}{(2\pi)^4} \Big[
ig_{\Upsilon BB} \epsilon_\Upsilon^\mu (ip_{1\mu}-ip_{2\mu}) \Big]
\Big[-ig_{BBV} (-ip_{1\nu}\nonumber\\ &&  -iq_\nu)
\epsilon_{\omega}^\nu\Big] \Big[-ig_{\chi_{b2}
BB}\epsilon_{\chi_{b2}}^{\alpha \beta} (-ip_{2\alpha}) (-iq_\beta)
\Big]\nonumber\\ && \times \frac{1}{p_1^2-m_B^2} \frac{1}{p_2^2
-m_B^2} \frac{1}{q^2 -m_B^2} \mathcal{F}^2(\Lambda),
\end{eqnarray}
\begin{eqnarray}
%%
%% B Bbar B^\ast
%%
\mathcal{M}_{(c-1)}^{B^\ast} &=& \int \frac{d^q}{(2\pi)^4} \Big[
ig_{\Upsilon BB} \epsilon_\Upsilon^\mu (ip_{1\mu}-ip_{2\mu}) \Big]
\Big[-2 f_{BB^\ast V} \varepsilon_{\lambda \nu \theta \phi}
\nonumber\\ && \times (ip_3^\lambda) \epsilon_\omega^\nu
(ip_1^\theta+iq^\theta) \Big] \Big[-g_{\chi_{b2} B^\ast B}
\varepsilon_{\alpha \tau \kappa \zeta} (ip_4^\kappa)
\epsilon_{\chi_{b2}}^{\alpha \beta} \nonumber\\ && \times
(-ip_{2\beta}) (-iq^\zeta) \Big] \frac{1}{p_1^2-m_B^2}
\frac{1}{p_2^2 -m_B^2} \nonumber\\ && \times \frac{-g^{\phi \tau}
+q^\phi q^\tau/m_{B^\ast}^2}{q^2 -m_{B^\ast}^2}
\mathcal{F}^2(\Lambda),
\end{eqnarray}
\begin{eqnarray}
%%
%% B Bstarbar B
%%
\mathcal{M}_{(c-2)}^{B} &=& \int \frac{d^q}{(2\pi)^4} \Big[
g_{\Upsilon B^\ast B} \varepsilon_{\rho \mu \tau \kappa}
(-ip_0^\rho) \epsilon_\Upsilon^\mu (-ip_2^\kappa +ip_1^\kappa)\Big]
\nonumber\\ && \times \Big[ -ig_{BBV}(-ip_{1\nu}-iq_\nu)
\epsilon_{\omega}^\nu \Big] \Big[-ig_{\chi_{b2} B^\ast B}
\varepsilon_{\alpha \theta \lambda
\phi} (ip_4^\lambda) \epsilon_{\chi_{b2}}^{\alpha \beta} \nonumber\\
&& \times (iq_\beta) (-ip_2^\phi)\Big] \frac{1}{p_1^2-m_B^2}
\frac{-g^{\tau \theta} +p_2^\tau p_2^\theta/m_{B^\ast}^2 }{p_2^2
-m_{B^\ast}^2} \nonumber\\
&& \times \frac{1}{q^2-m_B^2} \mathcal{F}^2(\Lambda),
\end{eqnarray}
\begin{eqnarray}
%%
%% B Bstarbar Bstar
%%
\mathcal{M}_{(c-2)}^{B^\ast} &=& \int \frac{d^q}{(2\pi)^4} \Big[
g_{\Upsilon B^\ast B} \varepsilon_{\rho \mu \tau \kappa}
(-ip_0^\rho) \epsilon_\Upsilon^\mu (-ip_2^\kappa +ip_1^\kappa)\Big]
\nonumber\\ && \times \Big[ -2 f_{BB^\ast V} \varepsilon_{\lambda
\nu \theta \phi} (ip_3^\lambda) \epsilon_\omega^\nu
(ip_1^\theta+iq^\theta) \Big] \Big[ g_{\chi_{b2} B^\ast B^\ast}
\epsilon_{\chi_{b2}}^{\alpha \beta}\Big] \nonumber\\ &&\times
\frac{1}{p_1^2-m_B^2} \frac{-g^{\kappa}_{\alpha} +p_2^\kappa
p_{2\alpha}/m_{B^\ast}^2 }{p_2^2
-m_{B^\ast}^2} \nonumber\\
&& \times \frac{-g_{\beta}^{\phi} +q_\beta
q^\phi/m_{B^\ast}^2}{q^2-m_{B^\ast}^2} \mathcal{F}^2(\Lambda),
\end{eqnarray}
\begin{eqnarray}
%%
%% Bstar Bbar B
%%
\mathcal{M}_{(c-3)}^{B} &=& \int \frac{d^q}{(2\pi)^4} \Big[
g_{\Upsilon B^\ast B} \varepsilon_{\rho \mu \tau \kappa}
(-ip_0^\rho) \epsilon_\Upsilon^\mu (ip_2^\kappa -ip_1^\kappa)\Big]
\nonumber\\ && \times \Big[ -2 f_{BB^\ast V} \varepsilon_{\lambda
\nu \theta \phi} (ip_3^\lambda) \epsilon_\omega^\nu
(-ip_1^\theta-iq^\theta) \Big] \nonumber\\
&&\times \Big[ g_{\chi_{b2} B B} \epsilon_{\chi_{b2}}^{\alpha
\beta}(-ip_{2\alpha}) (-iq_{\beta})\Big] \frac{-g^{\tau \phi}
+p_1^\tau p_1^\phi /m_{B^\ast}^2}{p_1^2-m_{B^\ast}^2} \nonumber\\
&&\times\frac{1 }{p_2^2 -m_{B}^2} \frac{1}{q^2-m_{B}^2}
\mathcal{F}^2(\Lambda),
\end{eqnarray}
\begin{eqnarray}
%%
%% Bstar Bbar BStar
%%
\mathcal{M}_{(c-3)}^{B^\ast} &=& \int \frac{d^q}{(2\pi)^4} \Big[
g_{\Upsilon B^\ast B} \varepsilon_{\rho \mu \tau \kappa}
(-ip_0^\rho) \epsilon_\Upsilon^\mu (ip_2^\kappa -ip_1^\kappa)\Big]
\nonumber\\ && \times \Big[ig_{B^\ast B^\ast V} (-ip_{1\nu}-iq_\nu)
\epsilon_\omega^\nu g_{\theta \phi} +4if_{B^\ast B^\ast V}
(ip_{3\phi} g_{\nu \theta}\nonumber\\ &&-ip_{3\theta} g_{\nu \phi})
\epsilon_{\omega}^\nu \Big] \Big[ -ig_{\chi_{b2} B^\ast B}
\varepsilon_{\alpha \zeta \lambda \delta } (ip_4^\lambda)
\epsilon_{\chi_{b2}}^{\alpha \beta} (iq_\beta) (-ip_2^\delta)
\Big]\nonumber\\
&&\times \frac{-g^{\tau \theta} +p_1^\tau p_1^\theta
/m_{B^\ast}^2}{p_1^2-m_{B^\ast}^2} \frac{1 }{p_2^2 -m_{B}^2}
\nonumber\\
&&\times \frac{-g^{\phi \zeta} +q^\phi
q^\zeta/m_{B^\ast}^2}{q^2-m_{B^\ast}^2} \mathcal{F}^2(\Lambda),
\end{eqnarray}
\begin{eqnarray}
%%
%% Bstar Bstarbar B
%%
\mathcal{M}_{(c-4)}^{B} &=& \int \frac{d^q}{(2\pi)^4}
\Big[ig_{\Upsilon B^\ast B^\ast} \epsilon_\Upsilon^\mu
(ip_{2\delta}g_{\mu \kappa}-ip_{1\kappa} g_{\mu \delta}
-(ip_{2\mu}\nonumber\\&&-ip_{1\mu})g_{\delta \kappa})\Big] \Big[-2
f_{B^\ast B V} \varepsilon_{\lambda \nu \gamma \phi} (ip_3^\lambda)
\epsilon_\omega^\nu (-ip_1^\gamma -iq^\gamma)\Big] \nonumber\\
&&\times \Big[-ig_{\chi_{b2} B^\ast B} \varepsilon_{\alpha \theta
\lambda \sigma} (ip_4^\lambda) \epsilon_{\chi_{b2}}^{\alpha \beta}
(iq_\beta) (-ip_2^\sigma)\Big]\nonumber\\ && \times \frac{-g^{\delta
\phi} +p_1^\delta p_1^\phi
/m_{B^\ast}^2}{p_1^2-m_{B^\ast}^2}\frac{-g^{\kappa \theta}
+p_2^\kappa p_2^\theta/m_{B^\ast}^2 }{p_2^2 -m_{B^\ast}^2}
\nonumber\\ && \times \frac{1}{q^2-m_{B}^2} \mathcal{F}^2(\Lambda),
\end{eqnarray}

\begin{eqnarray}
%%
%% Bstar Bstarbar Bstar
%%
\mathcal{M}_{(c-4)}^{B^\ast} &=& \int \frac{d^q}{(2\pi)^4}
\Big[ig_{\Upsilon B^\ast B^\ast} \epsilon_\Upsilon^\mu
(ip_{2\delta}g_{\mu \kappa}-ip_{1\kappa} g_{\mu \delta}
-(ip_{2\mu}\nonumber\\&&-ip_{1\mu})g_{\delta \kappa})\Big]
\Big[ig_{B^\ast B^\ast V} (-ip_{1\nu}-iq_\nu) \epsilon_\omega^\nu
g_{\theta \phi} +4if_{B^\ast B^\ast V} \nonumber\\ &&
\times(ip_{3\phi} g_{\nu \theta}-ip_{3\theta} g_{\nu \phi})
\epsilon_{\omega}^\nu \Big] \Big[g_{\chi_{b2} B^\ast B^\ast }
\epsilon_{\chi_{b2}}^{\alpha \beta}\Big] \nonumber\\ && \times
\frac{-g^{\delta \theta} +p_1^\delta p_1^\theta
/m_{B^\ast}^2}{p_1^2-m_{B^\ast}^2}\frac{-g^{\kappa}_{\alpha}
+p_2^\kappa p_{2\alpha}/m_{B^\ast}^2 }{p_2^2 -m_{B^\ast}^2} \nonumber\\
&& \times \frac{-g^{\phi}_{\beta} +q_\beta
q^\phi/m_{B^\ast}^2}{q^2-m_{B^\ast}^2} \mathcal{F}^2(\Lambda),
\end{eqnarray}
which correspond to Fig. \ref{Fig:Chib2J} (c-1)-(c-4), respectively.

\vfil

\section*{Acknowledgments}

This project is supported by the National Natural Science Foundation
of China under Grants No. 11222547, No. 11175073, No. 11035006 and
No. 11375240, the Ministry of Education of China (FANEDD under Grant
No. 200924, SRFDP under Grant No. 20120211110002, NCET, the
Fundamental Research Funds for the Central Universities); and the Fok
Ying Tung Education Foundation (No. 131006).

\end{document}